\documentclass[12pt,a4paper]{article}

\usepackage{ifthen} 
\newboolean{pdflatex}
\setboolean{pdflatex}{true} 

\newboolean{articletitles}
\setboolean{articletitles}{true} 

\newboolean{uprightparticles}
\setboolean{uprightparticles}{false} 

\newboolean{inbibliography}
\setboolean{inbibliography}{false} 

\usepackage{lineno}  
\usepackage{xspace} 
\usepackage{caption} 

\usepackage{graphicx}  
\usepackage{color}
\usepackage{colortbl}
\graphicspath{{./figs/}} 

\usepackage{amsmath} 
\usepackage{amssymb}
\usepackage{amsfonts}
\usepackage{upgreek} 

\newcommand*\patchAmsMathEnvironmentForLineno[1]{%
\expandafter\let\csname old#1\expandafter\endcsname\csname #1\endcsname
\expandafter\let\csname oldend#1\expandafter\endcsname\csname
end#1\endcsname
 \renewenvironment{#1}%
   {\linenomath\csname old#1\endcsname}%
   {\csname oldend#1\endcsname\endlinenomath}%
}
\newcommand*\patchBothAmsMathEnvironmentsForLineno[1]{%
  \patchAmsMathEnvironmentForLineno{#1}%
  \patchAmsMathEnvironmentForLineno{#1*}%
}
\AtBeginDocument{%
\patchBothAmsMathEnvironmentsForLineno{equation}%
\patchBothAmsMathEnvironmentsForLineno{align}%
\patchBothAmsMathEnvironmentsForLineno{flalign}%
\patchBothAmsMathEnvironmentsForLineno{alignat}%
\patchBothAmsMathEnvironmentsForLineno{gather}%
\patchBothAmsMathEnvironmentsForLineno{multline}%
}

\usepackage{hyperref}    
\usepackage[all]{hypcap} 

\def\lhcb {\mbox{LHCb}\xspace}

\ifthenelse{\boolean{uprightparticles}}%
{

 \def\Ppi         {\ensuremath{\uppi}\xspace}

 \def\Ppsi        {\ensuremath{\uppsi}\xspace}

 \def\PDelta      {\ensuremath{\Delta}\xspace}                 
 \def\PXi      {\ensuremath{\Xi}\xspace}                 
 \def\PLambda      {\ensuremath{\Lambda}\xspace}                 
 \def\PSigma      {\ensuremath{\Sigma}\xspace}                 
 \def\POmega      {\ensuremath{\Omega}\xspace}                 
 \def\PUpsilon      {\ensuremath{\Upsilon}\xspace}                 
 
 \def\PB      {\ensuremath{\mathrm{B}}\xspace}                 
                  
 \def\PD      {\ensuremath{\mathrm{D}}\xspace}

 \def\PJ      {\ensuremath{\mathrm{J}}\xspace}                 
 \def\PK      {\ensuremath{\mathrm{K}}\xspace}

 \def\Pb      {\ensuremath{\mathrm{b}}\xspace}                 
 \def\Pc      {\ensuremath{\mathrm{c}}\xspace}

 \def\Pi      {\ensuremath{\mathrm{i}}\xspace}

}
{

 \def\Ppi         {\ensuremath{\pi}\xspace}

 \def\Ppsi        {\ensuremath{\psi}\xspace}                 
                  
 \mathchardef\PDelta="7101
 \mathchardef\PXi="7104
 \mathchardef\PLambda="7103
 \mathchardef\PSigma="7106
 \mathchardef\POmega="710A
 \mathchardef\PUpsilon="7107
                  
 \def\PB      {\ensuremath{B}\xspace}                 
                  
 \def\PD      {\ensuremath{D}\xspace}

 \def\PJ      {\ensuremath{J}\xspace}                 
 \def\PK      {\ensuremath{K}\xspace}

 \def\Pb      {\ensuremath{b}\xspace}                 
 \def\Pc      {\ensuremath{c}\xspace}

 \def\Pi      {\ensuremath{i}\xspace}

}

\def\cquark    {{\ensuremath{\Pc}}\xspace}

\def\bquark    {{\ensuremath{\Pb}}\xspace}

\def\pion   {{\ensuremath{\Ppi}}\xspace}

\def\pip    {{\ensuremath{\pion^+}}\xspace}
\def\pim    {{\ensuremath{\pion^-}}\xspace}

\def\kaon    {{\ensuremath{\PK}}\xspace}
  \def\Kbar    {{\kern 0.2em\overline{\kern -0.2em \PK}{}}\xspace}

\def\Kp      {{\ensuremath{\kaon^+}}\xspace}
\def\Km      {{\ensuremath{\kaon^-}}\xspace}

\def\Kstar   {{\ensuremath{\kaon^*}}\xspace}

  \def\Dbar    {{\kern 0.2em\overline{\kern -0.2em \PD}{}}\xspace}
\def\D       {{\ensuremath{\PD}}\xspace}

\def\Dz      {{\ensuremath{\D^0}}\xspace}

\def\Dp      {{\ensuremath{\D^+}}\xspace}

\def\B       {{\ensuremath{\PB}}\xspace}
\def\Bbar    {{\ensuremath{\kern 0.18em\overline{\kern -0.18em \PB}{}}}\xspace}

\def\Bz      {{\ensuremath{\B^0}}\xspace}

\def\jpsi     {{\ensuremath{{\PJ\mskip -3mu/\mskip -2mu\Ppsi\mskip 2mu}}}\xspace}

  \def\Y#1S{\ensuremath{\PUpsilon{(#1S)}}\xspace}

\def\Xires       {{\ensuremath{\PXi}}\xspace}

\def\Lz          {{\ensuremath{\PLambda}}\xspace}
\def\Lbar        {{\ensuremath{\kern 0.1em\overline{\kern -0.1em\PLambda}}}\xspace}

\def\Sigmares    {{\ensuremath{\PSigma}}\xspace}

\def\Lb      {{\ensuremath{\Lz^0_\bquark}}\xspace}

\def\Lc      {{\ensuremath{\Lz^+_\cquark}}\xspace}
\def\Xic      {{\ensuremath{\PXi^+_\cquark}}\xspace}

\def\Xibz      {{\ensuremath{\Xires^0_\bquark}}\xspace}
\def\Xibm      {{\ensuremath{\Xires^-_\bquark}}\xspace}
\def\Ombm      {{\ensuremath{\POmega_{\bquark}^{-}}}\xspace}

\def\to                 {\ensuremath{\rightarrow}\xspace}

\def\AT#1     {\ensuremath{A_{\mathrm{T}}^{#1}}\xspace}           

\def\C#1      {\ensuremath{\mathcal{C}_{#1}}\xspace}                       
\def\Cp#1     {\ensuremath{\mathcal{C}_{#1}^{'}}\xspace}                    
\def\Ceff#1   {\ensuremath{\mathcal{C}_{#1}^{\mathrm{(eff)}}}\xspace}        
\def\Cpeff#1  {\ensuremath{\mathcal{C}_{#1}^{'\mathrm{(eff)}}}\xspace}       
\def\Ope#1    {\ensuremath{\mathcal{O}_{#1}}\xspace}                       
\def\Opep#1   {\ensuremath{\mathcal{O}_{#1}^{'}}\xspace}                    

\newcommand{\tev}{\ifthenelse{\boolean{inbibliography}}{\ensuremath{~T\kern -0.05em eV}\xspace}{\ensuremath{\mathrm{\,Te\kern -0.1em V}}}\xspace}
\newcommand{\gev}{\ensuremath{\mathrm{\,Ge\kern -0.1em V}}\xspace}
\newcommand{\mev}{\ensuremath{\mathrm{\,Me\kern -0.1em V}}\xspace}
\newcommand{\kev}{\ensuremath{\mathrm{\,ke\kern -0.1em V}}\xspace}
\newcommand{\ev}{\ensuremath{\mathrm{\,e\kern -0.1em V}}\xspace}
\newcommand{\gevc}{\ensuremath{{\mathrm{\,Ge\kern -0.1em V\!/}c}}\xspace}
\newcommand{\mevc}{\ensuremath{{\mathrm{\,Me\kern -0.1em V\!/}c}}\xspace}
\newcommand{\gevcc}{\ensuremath{{\mathrm{\,Ge\kern -0.1em V\!/}c^2}}\xspace}
\newcommand{\gevgevcccc}{\ensuremath{{\mathrm{\,Ge\kern -0.1em V^2\!/}c^4}}\xspace}
\newcommand{\mevcc}{\ensuremath{{\mathrm{\,Me\kern -0.1em V\!/}c^2}}\xspace}

\def\mum  {\ensuremath{{\,\upmu\rm m}}\xspace}

\def\invfb   {\ensuremath{\mbox{\,fb}^{-1}}\xspace}

\def\ps   {\ensuremath{{\rm \,ps}}\xspace}

\newcommand{\chisq}{\ensuremath{\chi^2}\xspace}

\newcommand{\chisqip}{\ensuremath{\chi^2_{\rm IP}}\xspace}
\newcommand{\chisqvs}{\ensuremath{\chi^2_{\rm VS}}\xspace}
\newcommand{\chisqvtx}{\ensuremath{\chi^2_{\rm vtx}}\xspace}

\def\gsim{{~\raise.15em\hbox{$>$}\kern-.85em
          \lower.35em\hbox{$\sim$}~}\xspace}
\def\lsim{{~\raise.15em\hbox{$<$}\kern-.85em
          \lower.35em\hbox{$\sim$}~}\xspace}

\def\pt         {\mbox{$p_{\rm T}$}\xspace}

\def\evtgen     {\mbox{\textsc{EvtGen}}\xspace}

\def\gauss      {\mbox{\textsc{Gauss}}\xspace}
\def\geant      {\mbox{\textsc{Geant4}}\xspace}

\def\photos     {\mbox{\textsc{Photos}}\xspace}

\def\pythia     {\mbox{\textsc{Pythia}}\xspace}

\def\tell1  {TELL1\xspace}
\def\ukl1   {UKL1\xspace}

\def\br{{\cal{B}}}

\def\eff{\epsilon}

\def\rstat{{\rm stat}}
\def\rsyst{{\rm syst}}

\def\eff{\epsilon}

\usepackage{cite} 
\usepackage{mciteplus}

\usepackage{longtable} 

\begin{document}

\renewcommand{\thefootnote}{\fnsymbol{footnote}}
\setcounter{footnote}{1}


\begin{titlepage}
\pagenumbering{roman}

\vspace*{-1.5cm}
\centerline{\large EUROPEAN ORGANIZATION FOR NUCLEAR RESEARCH (CERN)}
\vspace*{1.5cm}
\hspace*{-0.5cm}
\begin{tabular*}{\linewidth}{lc@{\extracolsep{\fill}}r}
\ifthenelse{\boolean{pdflatex}}
{\vspace*{-2.7cm}\mbox{\!\!\!\includegraphics[width=.14\textwidth]{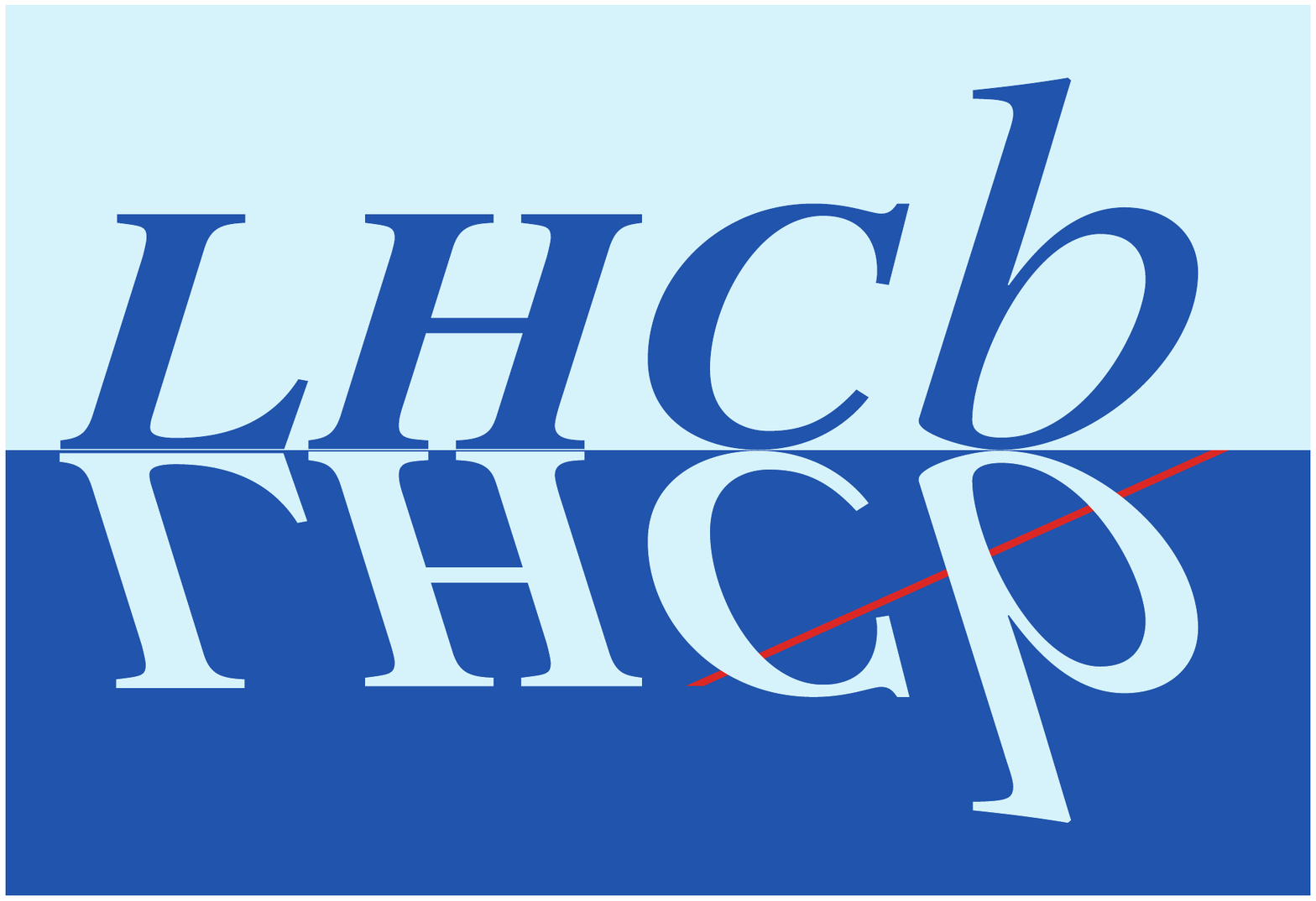}} & &}%
{\vspace*{-1.2cm}\mbox{\!\!\!\includegraphics[width=.12\textwidth]{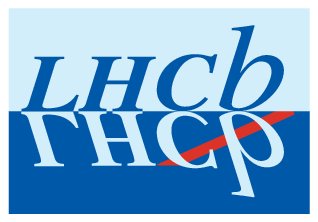}} & &}%
\\
 & & CERN-PH-EP-2014-107 \\  
 & & LHCb-PAPER-2014-021 \\  
 & & 16 July 2014 \\ 
 & & \\
\end{tabular*}

\vspace*{0.0cm}

{\bf\boldmath\huge
\begin{center}
  Precision measurement of the mass and lifetime of the $\Xibz$ baryon
\end{center}
}

\vspace*{0.5cm}

\begin{center}
The LHCb collaboration\footnote{Authors are listed on the following pages.}
\end{center}

\vspace{\fill}

\begin{abstract}
  \noindent
Using a proton-proton collision data sample corresponding to an integrated luminosity of 3\invfb collected by LHCb at
center-of-mass energies of 7 and 8~\tev, about 3800 $\Xibz\to\Xic\pim$, $\Xic\to p\Km\pip$  signal decays are reconstructed.
From this sample, the first measurement of the $\Xibz$ baryon lifetime is made, relative to that of the
$\Lb$ baryon. The mass differences $M(\Xibz)-M(\Lb)$ and $M(\Xic)-M(\Lc)$ are also measured 
with precision more than four times better than the current world averages. The resulting values are
\begin{align*}
\frac{\tau_{\Xibz}}{\tau_{\Lb}} &= 1.006\pm0.018\pm0.010, \\
M(\Xibz) - M(\Lb) &= 172.44\pm0.39\pm0.17~\mevcc, \\
M(\Xic) - M(\Lc) &= 181.51\pm0.14\pm0.10~\mevcc, 
\end{align*}
\noindent where the first uncertainty is statistical and the second is systematic.
The relative rate of $\Xibz$ to $\Lb$ baryon production is measured to be
\begin{align*}
\frac{f_{\Xibz}}{f_{\Lb}}\cdot\frac{\br(\Xibz\to\Xic\pim)}{\br(\Lb\to\Lc\pim)}\cdot\frac{\br(\Xic\to p\Km\pip)}{\br(\Lc\to p\Km\pip)} = (1.88\pm0.04\pm0.03)\times10^{-2}, 
\end{align*}
\noindent where the first factor is the ratio of fragmentation fractions, $b\to\Xibz$ relative to $b\to\Lb$.
Relative production rates as functions of transverse momentum and pseudorapidity are also presented.
\end{abstract}

\vspace*{0.0cm}

\begin{center}
  Submitted to Phys.~Rev.~Lett. 
\end{center}

\vspace{\fill}

{\footnotesize 
\centerline{\copyright~CERN on behalf of the \lhcb collaboration, license \href{http://creativecommons.org/licenses/by/3.0/}{CC-BY-3.0}.}}
\vspace*{0.2mm}

\end{titlepage}


\newpage
\setcounter{page}{2}
\mbox{~}
\newpage

\centerline{\large\bf LHCb collaboration}
\begin{flushleft}
\small
R.~Aaij$^{41}$, 
B.~Adeva$^{37}$, 
M.~Adinolfi$^{46}$, 
A.~Affolder$^{52}$, 
Z.~Ajaltouni$^{5}$, 
S.~Akar$^{6}$, 
J.~Albrecht$^{9}$, 
F.~Alessio$^{38}$, 
M.~Alexander$^{51}$, 
S.~Ali$^{41}$
% ...

{\footnotesize \it
% ...
$ ^{5}$Clermont Universit\'{e}, Universit\'{e} Blaise Pascal, CNRS/IN2P3, LPC, Clermont-Ferrand, France\\
$ ^{6}$CPPM, Aix-Marseille Universit\'{e}, CNRS/IN2P3, Marseille, France\\
% ...
$ ^{9}$Fakult\"{a}t Physik, Technische Universit\"{a}t Dortmund, Dortmund, Germany\\
% ...
$ ^{37}$Universidad de Santiago de Compostela, Santiago de Compostela, Spain\\
$ ^{38}$European Organization for Nuclear Research (CERN), Geneva, Switzerland\\
% ...
$ ^{41}$Nikhef National Institute for Subatomic Physics, Amsterdam, The Netherlands\\
% ...
$ ^{46}$H.H. Wills Physics Laboratory, University of Bristol, Bristol, United Kingdom\\
% ...
$ ^{51}$School of Physics and Astronomy, University of Glasgow, Glasgow, United Kingdom\\
$ ^{52}$Oliver Lodge Laboratory, University of Liverpool, Liverpool, United Kingdom\\
% ...
$ ^{p}$LIFAELS, La Salle, Universitat Ramon Llull, Barcelona, Spain\\
% ...
}
\end{flushleft}

\cleardoublepage


\renewcommand{\thefootnote}{\arabic{footnote}}
\setcounter{footnote}{0}



\pagestyle{plain} 
\setcounter{page}{1}
\pagenumbering{arabic}


%

  Over the last two decades great progress has been made in understanding the nature of hadrons containing
beauty quarks. A number of theoretical tools have been developed to 
describe their decays. One of them, the heavy quark expansion (HQE)
~\cite{Khoze:1983yp,Bigi:1991ir,Bigi:1992su,Blok:1992hw,Blok:1992he,Neubert:1997gu,Uraltsev:1998bk,Bigi:1995jr},
expresses the decay widths as an expansion in powers of $\Lambda_{\rm QCD}/m_b$, where $\Lambda_{\rm QCD}$ 
is the energy scale at which the strong coupling constant becomes large, and $m_b$ is the $b$-quark mass.
At leading order in the HQE, all weakly decaying $b$ hadrons (excluding those containing charm quarks) 
have the same lifetime, and differences enter only
at order $(\Lambda_{\rm QCD}/m_b)^2$. In the baryon sector, one expects for the lifetimes
$\tau(\Xibz)\approx\tau(\Lb)$~\cite{Bigi:1995jr} and 
$\tau(\Xibz)/\tau(\Xibm)=0.95\pm0.06$~\cite{Lenz:2008xt,Lenz:LHQE2014}.
Precise measurements of the $\Xibz$ and $\Xibm$ lifetimes would
put bounds on the magnitude of the higher order terms in the HQE.
A number of approaches exist to predict the $b$-baryon
masses~\cite{oai:arXiv.org:hep-ph/0504112,oai:arXiv.org:hep-ph/0203253,oai:arXiv.org:0710.0123,oai:arXiv.org:0712.0406,oai:arXiv.org:hep-ph/9502251,Karliner:2008sv,oai:arXiv.org:0806.4951,Ghalenovi,Zhang:2008pm}. As predictions for the masses span a large range, more precise mass measurements will help to refine these models.

Hadron collider experiments have collected large samples of $b$-baryon decays, which have enabled
increasingly precise measurements of their masses and 
lifetimes~\cite{LHCb-PAPER-2014-003,LHCb-PAPER-2013-065,LHCb-PAPER-2014-002,LHCb-PAPER-2014-010,Aad:2012bpa,Chatrchyan:2013sxa}.
These advances include 1\% precision on the lifetime of the $\Lb$ baryon~\cite{LHCb-PAPER-2014-003} and 
0.3\mevcc uncertainty on its mass~\cite{LHCb-PAPER-2014-002}. Progress has also been made on
improving the precision on the masses of the $\Sigmares_b^{\pm}$~\cite{CDF:2011ac},
$\Xibz$~\cite{LHCb-PAPER-2013-056,Aaltonen:2011wd,Aaltonen:2014wfa}, 
$\Xibm$~\cite{LHCb-PAPER-2012-048,CDF:2011ac} 
and $\Ombm$~\cite{LHCb-PAPER-2012-048,CDF:2011ac} baryons. 
The strange-beauty baryon measurements are still limited by small sample sizes owing to their 
low production rates, and either low detection efficiency or small branching fractions. 

In this Letter, we present the first measurement of the $\Xibz$ lifetime and report the most precise
measurement of its mass, using a sample of about 3800 $\Xibz\to\Xic\pim$, $\Xic\to p\Km\pip$ signal decays.
Unless otherwise noted, charge conjugate processes are implied throughout. The $\Lb\to\Lc\pim$, $\Lc\to p\Km\pip$
decay is used for normalization, as it has the same final state, and is kinematically very similar. The
ratio of $\Xibz$ to $\Lb$ baryon production rates, and its dependence on pseudorapidity, $\eta$, and
transverse momentum, \pt, are also presented. We also use the $\Xic\to p\Km\pip$ and $\Lc\to p\Km\pip$ signals
to make the most precise measurement of the $\Xic$ mass to date. In what follows,
we use $X_b$ ($X_c$) to refer to either a $\Xibz$ ($\Xic$) or $\Lb$ ($\Lc$) baryon.

The measurements use proton-proton ($pp$) collision data samples collected by the LHCb experiment
corresponding  to an integrated luminosity of 3~\invfb, of which 1\invfb was recorded at 
a center-of-mass energy of 7\tev and 2\invfb at 8\tev.  
The \lhcb detector~\cite{Alves:2008zz} is a single-arm forward
spectrometer covering the \mbox{pseudorapidity} range $2<\eta <5$,
designed for the study of particles containing \bquark or \cquark
quarks. The detector includes a high-precision tracking system
that provides a momentum measurement with precision of about 0.5\% from
2$-$100~\gevc and impact parameter (IP) resolution of 20\mum for
particles with large \pt. Ring-imaging Cherenkov detectors~\cite{LHCb-DP-2012-003}
are used to distinguish charged hadrons. Photon, electron and
hadron candidates are identified using a calorimeter system, followed by a set
of detectors to identify muons~\cite{LHCb-DP-2012-002}.

The trigger~\cite{LHCb-DP-2012-004} consists of a
hardware stage, based on information from the calorimeter and muon
systems, followed by a software stage, which applies a full event
reconstruction~\cite{LHCb-DP-2012-004,BBDT}. 
About 57\% of the recorded $X_b$ events are triggered at the hardware 
level by one or more of the final state particles in the signal $X_b$ decay. 
The remaining 43\% are triggered only on other activity in the event.
We refer to these two classes of events as triggered on signal (TOS) and
triggered independently of signal (TIS).
The software trigger requires a two-, three- or four-track
secondary vertex with a large sum of the transverse momentum of
the particles and a significant displacement from the primary $pp$
interaction vertices~(PVs). At least one particle should have $\pt>1.7\gevc$ and \chisqip with respect to any
primary interaction greater than 16, where \chisqip is defined as the
difference in \chisq of a given PV fitted with and
without the considered particle included. The signal candidates 
are required to pass a multivariate software trigger
selection algorithm~\cite{BBDT}. 

Proton-proton collisions are simulated using
\pythia~\cite{Sjostrand:2006za,*Sjostrand:2007gs} with a specific \lhcb
configuration~\cite{LHCb-PROC-2010-056}.  Decays of hadronic particles
are described by \evtgen~\cite{Lange:2001uf}, in which final state
radiation is generated using \photos~\cite{Golonka:2005pn}. The
interaction of the generated particles with the detector and its
response are implemented using the \geant toolkit~\cite{Allison:2006ve, *Agostinelli:2002hh} as described in
Ref.~\cite{LHCb-PROC-2011-006}.

Candidate $X_b$ decays are reconstructed by combining in a kinematic fit selected $X_c\to p\Km\pip$ candidates
with a $\pim$ candidate (referred to as the bachelor). Each $X_b$ candidate is associated to the 
PV with the smallest $\chisqip$. The $X_c$ daughters are required to
have $\pt>100$\mevc, and the bachelor pion is required to have \pt$>$500\mevc.
To improve the signal purity, all four final state particles are required to be significantly 
displaced from the PV and pass particle identification (PID) requirements.
The PID requirements on the $X_c$ daughter particles have an efficiency of
74\%, while reducing the combinatorial background by a factor of four. The PID requirements on the
bachelor pion are 98\% efficient, and remove about 60\% of the cross-feed from $X_b\to X_c\Km$ decays. 
Cross-feed from misidentified $D^{+}_{(s)}\to\Kp\Km\pip$,
$D^{*+}\to\Dz(\Kp\Km)\pip$, and $\Dp\to\Km\pip\pip$ decays is removed by requiring either the mass under these 
alternate decay hypotheses to be inconsistent with the known $D^{(*)+}_{(s)}$ masses~\cite{PDG2012}, or that the
candidate satisfy more stringent PID requirements. The efficiency of these vetoes is about 98\% and they
reject 28\% of the background. The $X_c$ candidate is required to be within 20\mevcc of the nominal 
$X_c$ mass~\cite{PDG2012}.

To further improve the signal-to-background ratio, a boosted decision tree (BDT)~\cite{Breiman,AdaBoost} algorithm
using eight input variables is employed. Three variables from the $X_b$ candidate are used, $\chisqip$, 
the vertex fit $\chisqvtx$, and the $\chisqvs$, which
is the increase in $\chi^2$ of the PV fit when the $X_b$ is forced to have zero lifetime
relative to the nominal fit. For the $X_c$ baryon, we use the $\chisqip$, and amongst its daughters,
we take the minimum $\pt$, the smallest $\chisqip$, and the largest distance between any pair of
daughter particles. Lastly, the $\chisqip$ of the bachelor $\pim$ is used. 
The BDT is trained using
simulated signal decays to represent the signal and candidates from the high $X_b$ mass region (beyond the
fit region) to describe the background distributions. A selection is applied that 
provides 97\% signal efficiency while rejecting about 50\% of the combinatorial background with respect
to all previously applied selections.

For each $X_b$ candidate, the mass is recomputed using vertex constraints to improve the momentum resolution;
$X_c$ mass constraints are not used since the $\Xic$ mass is not known to sufficient precision.
The resulting $X_b$ mass spectra are simultaneously fitted to the sum of a signal component and three background contributions.
The $X_b$ signal shape is parameterized as the sum of two Crystal Ball (CB) functions~\cite{Skwarnicki:1986xj},
with a common mean. The shape parameters are freely varied in the fit to data. The $\Lb$ and $\Xibz$ signal shape
parameters are common except for their means and widths. The $\Xibz$ widths are fixed to be 0.6\% 
larger than those for the $\Lb$, based on simulation.

The main background sources are misidentified $X_b\to X_c\Km$ decays, partially reconstructed $X_b\to X_c\rho^-$
and $\Lb\to\Sigmares_c^+\pim$ decays, and combinatorial background. The $X_b\to X_c\Km$ background shape is
obtained from simulated decays that are weighted according to PID misidentification rates
obtained from $D^{*+}\to\Dz(\Km\pip)\pip$ calibration data. The $X_b\to X_c\Km$ yield is fixed to be
3.1\% of the $X_b\to X_c\pim$ signal yield, which is the product of the misidentification rate
of 42\% and the ratio of branching fractions, $\br(\Lb\to\Lc\Km)/\br(\Lb\to\Lc\pim)=0.0731\pm0.0023$~\cite{LHCb-PAPER-2013-056}. 
The assumed equality of this ratio for $\Xibz$ and $\Lb$ is considered as a source of systematic uncertainty.
The partially reconstructed backgrounds are modeled empirically using an ARGUS~\cite{Albrecht:1994tb} function, convolved with a Gaussian shape;
all of its shape parameters are freely varied in the fit. The combinatorial background shape is
described using an exponential function with a freely varied shape parameter.

\begin{figure}[tb]
\centering
\includegraphics[width=0.48\textwidth]{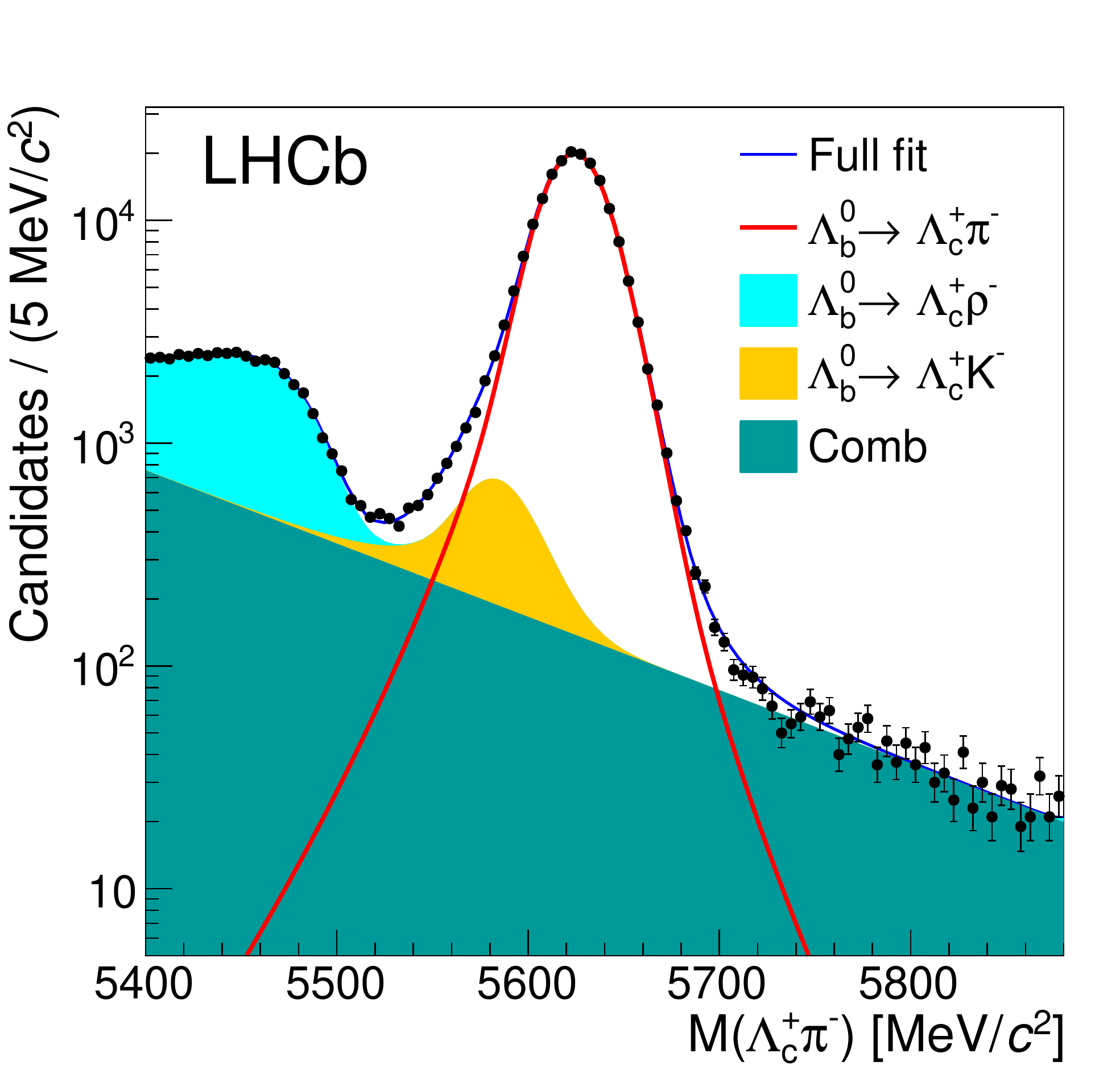}
\includegraphics[width=0.48\textwidth]{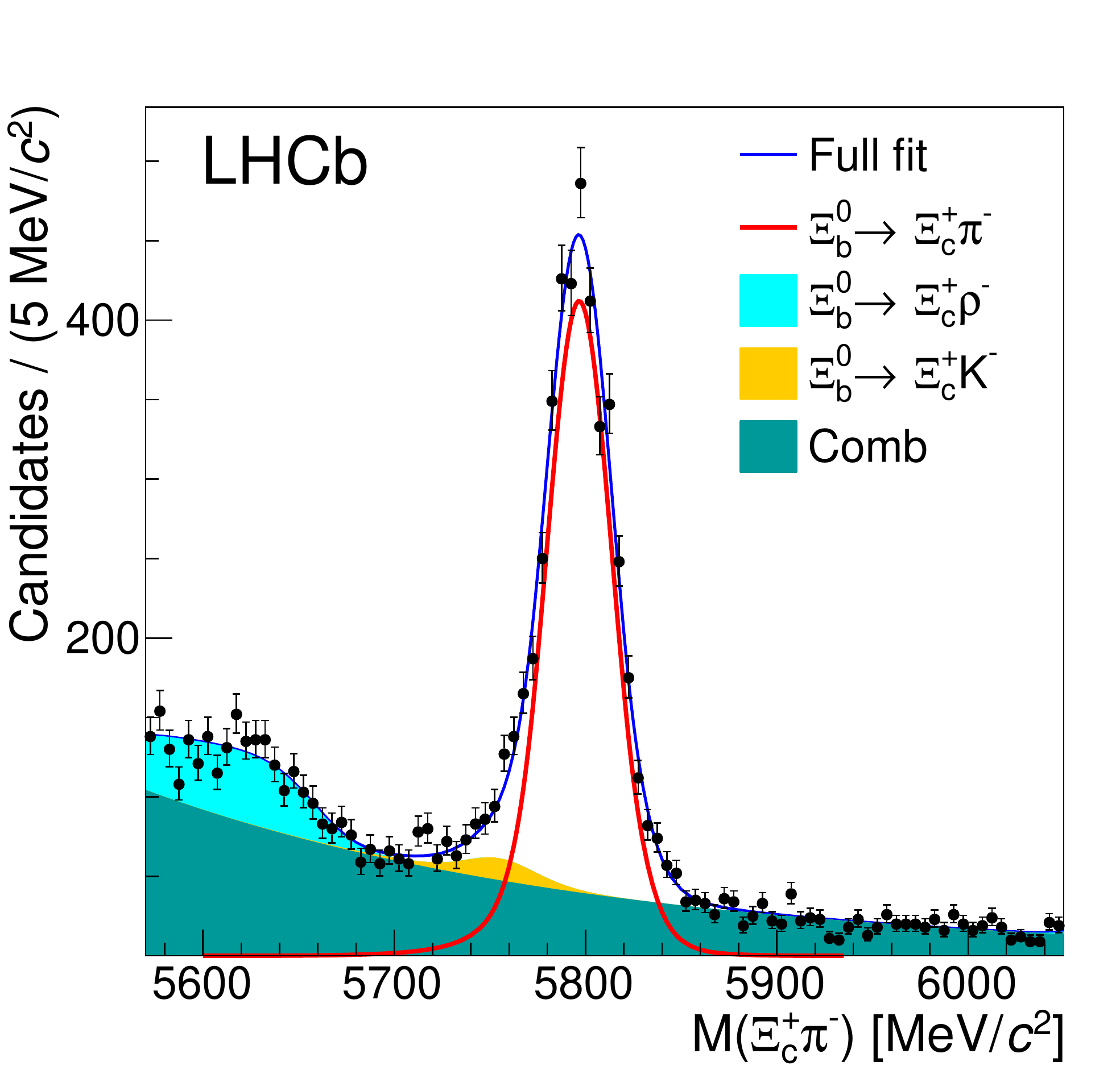}
\caption{\small{Invariant mass spectrum for (left) $\Lb\to\Lc\pim$ and (right) $\Xibz\to\Xic\pim$
candidates along with the projections of the fit.}}
\label{fig:XbMassFits}
\end{figure}

The results of the simultaneous binned extended maximum likelihood fits are shown in Fig.~\ref{fig:XbMassFits}.
Peaking backgrounds from charmless final states are investigated using the $X_c$ sidebands and are found to be negligible.
We observe $(180.5\pm0.5)\times10^{3}$ $\Lb\to\Lc\pim$ and $3775\pm71$ $\Xibz\to\Xic\pim$ signal decays.
The mass difference is determined to be
\begin{align*}
\Delta M_{X_b}\equiv M(\Xibz)-M(\Lb) = 172.44\pm0.39\,(\rstat) \mevcc.
\end{align*}
The data are also used to make the first determination of the relative lifetime $\tau(\Xibz)/\tau(\Lb)$.
This is performed by fitting the efficiency-corrected ratio of yields, $N_{\rm cor}(\Xibz)/N_{\rm cor}(\Lb)$,
as a function of decay time to an exponential function, $e^{\beta t}$. The fitted value of $\beta$ thus
determines $1/\tau_{\Lb} - 1/\tau_{\Xibz}$. Since the $\Lb$ lifetime is known to high precision, 
$\tau(\Xibz)$ is readily obtained.
The data are binned in 0.5~ps bins from $0-6$~ps, and 1~ps bins from 7 to 9~ps. The same fit as described
above for the full sample is used to fit the mass spectra in each time bin. The signal
and partially-reconstructed background shapes are fixed to the values from the fit to the
full data sample, since they do not change with decay time, but the combinatorial background shape
is freely varied in each time bin fit.

The measured yield ratio in each time bin is corrected by the relative efficiency, $\eff(\Lb)/\eff(\Xibz)$,
as obtained from simulated decays. This ratio is consistent with a constant value of about 0.93, except for the $0.0-0.5$~ps bin, 
which has a value of about 0.7. This lower value is expected due to the differing lifetimes, $\tau(\Xic)\approx0.45~{\rm ps}\gg\tau(\Lc)\approx0.2~{\rm ps}$,
and the $\chisqip$ requirements in the trigger and offline selections. 
The 7\% overall lower efficiency for the $\Lb$ mode is due to the larger momenta of the daughters in the $\Xibz$ decay.

\begin{figure}[tb]
\centering
\includegraphics[width=1.0\textwidth]{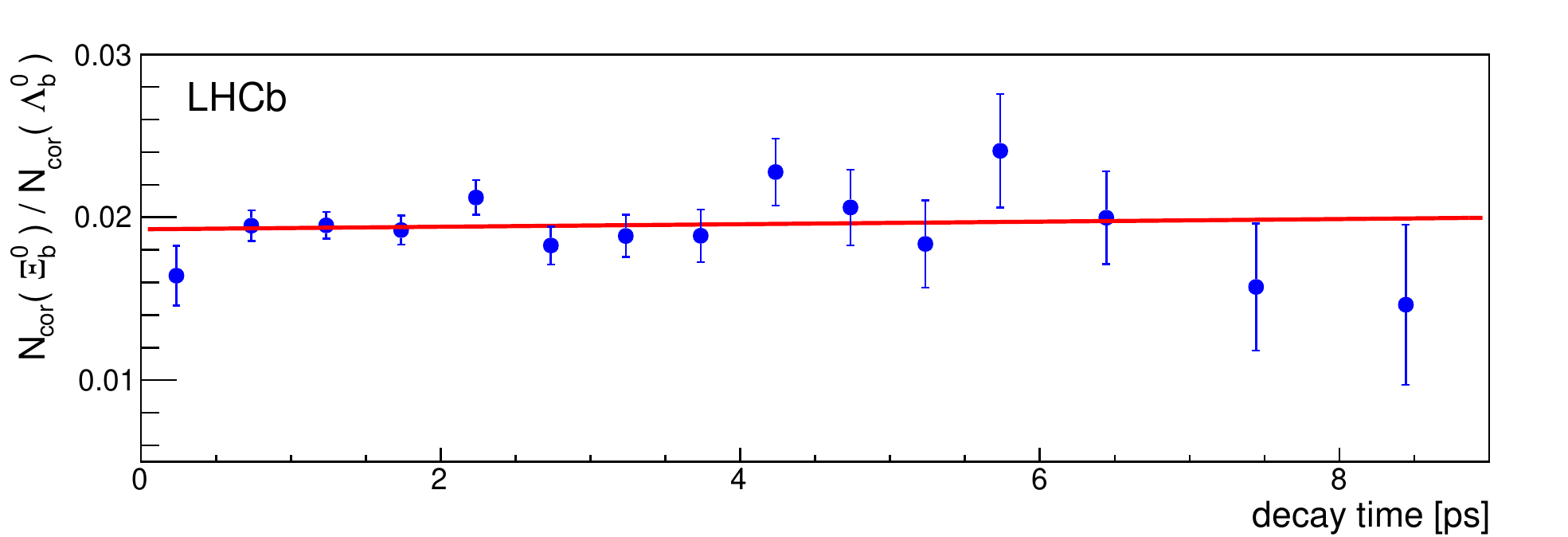}
\caption{\small{Efficiency-corrected yield ratio of $\Xibz\to\Xic\pim$ relative to $\Lb\to\Lc\pim$ decays in bins of decay time. 
A fit using an exponential function is shown. The uncertainties are statistical only.}}
\label{fig:CorrYieldRatio}
\end{figure}

The efficiency-corrected yield ratio is shown in Fig.~\ref{fig:CorrYieldRatio}, along with the fit to an exponential function.
The points are placed at the weighted average time value within each bin, assuming an 
exponential distribution with lifetime equal to $\tau(\Lb)$. The bias due to this assumption is negligible.
From the fit, we find $\beta = (0.40\pm1.21)\times10^{-2}~{\rm ps}^{-1}$.
Using the measured $\Lb$ lifetime from \lhcb of 
$1.468\pm0.009\pm0.008$~ps~\cite{LHCb-PAPER-2014-003}, we obtain
\begin{align*}
\frac{\tau_{\Xibz}}{\tau_{\Lb}} = \frac{1}{1-\beta\tau_{\Lb}} = 1.006\pm0.018\,(\rstat),
\end{align*}
\noindent consistent with equal lifetimes of the $\Xibz$ and $\Lb$ baryons. 

\begin{figure}[tb]
\centering
\includegraphics[width=0.48\textwidth]{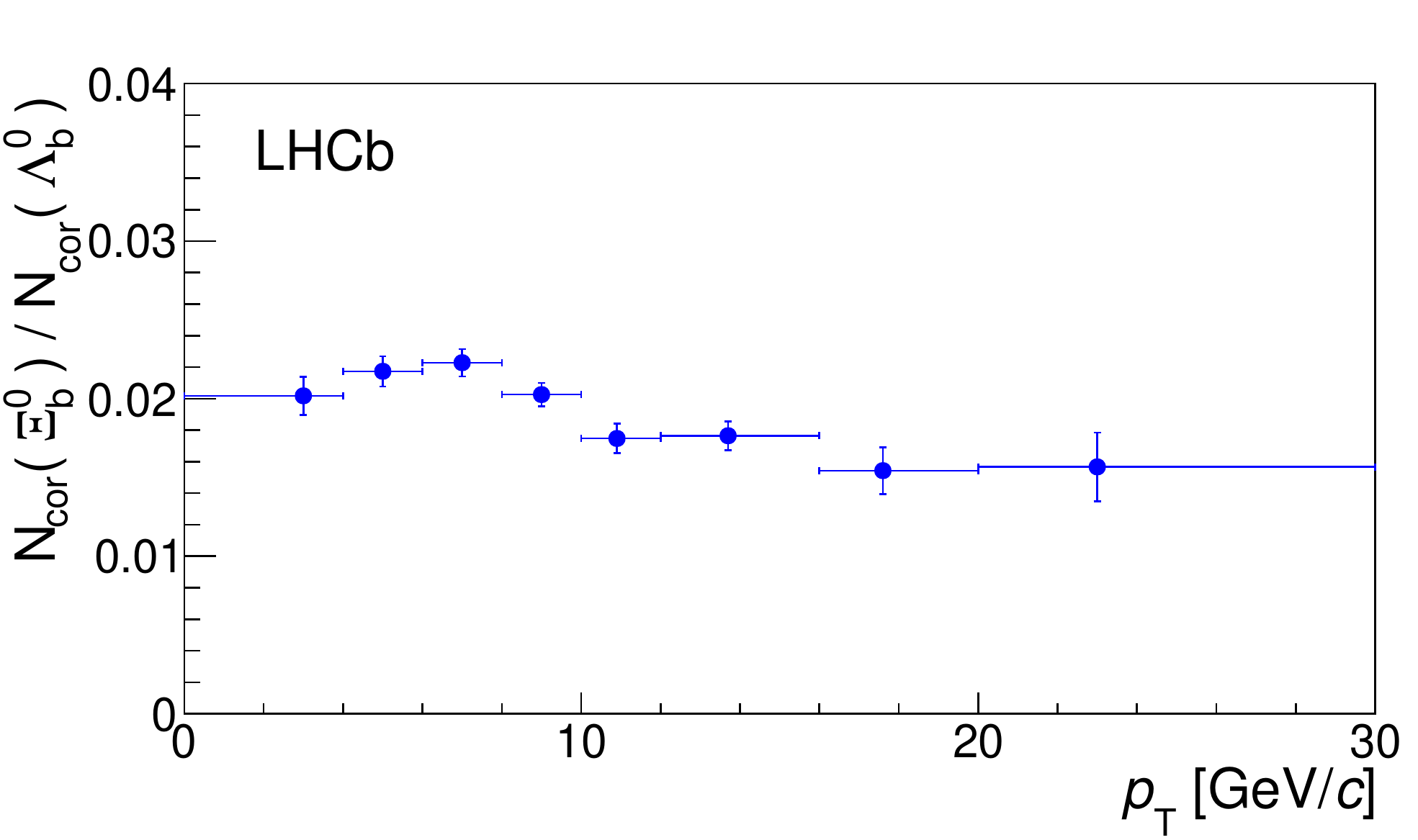}
\includegraphics[width=0.48\textwidth]{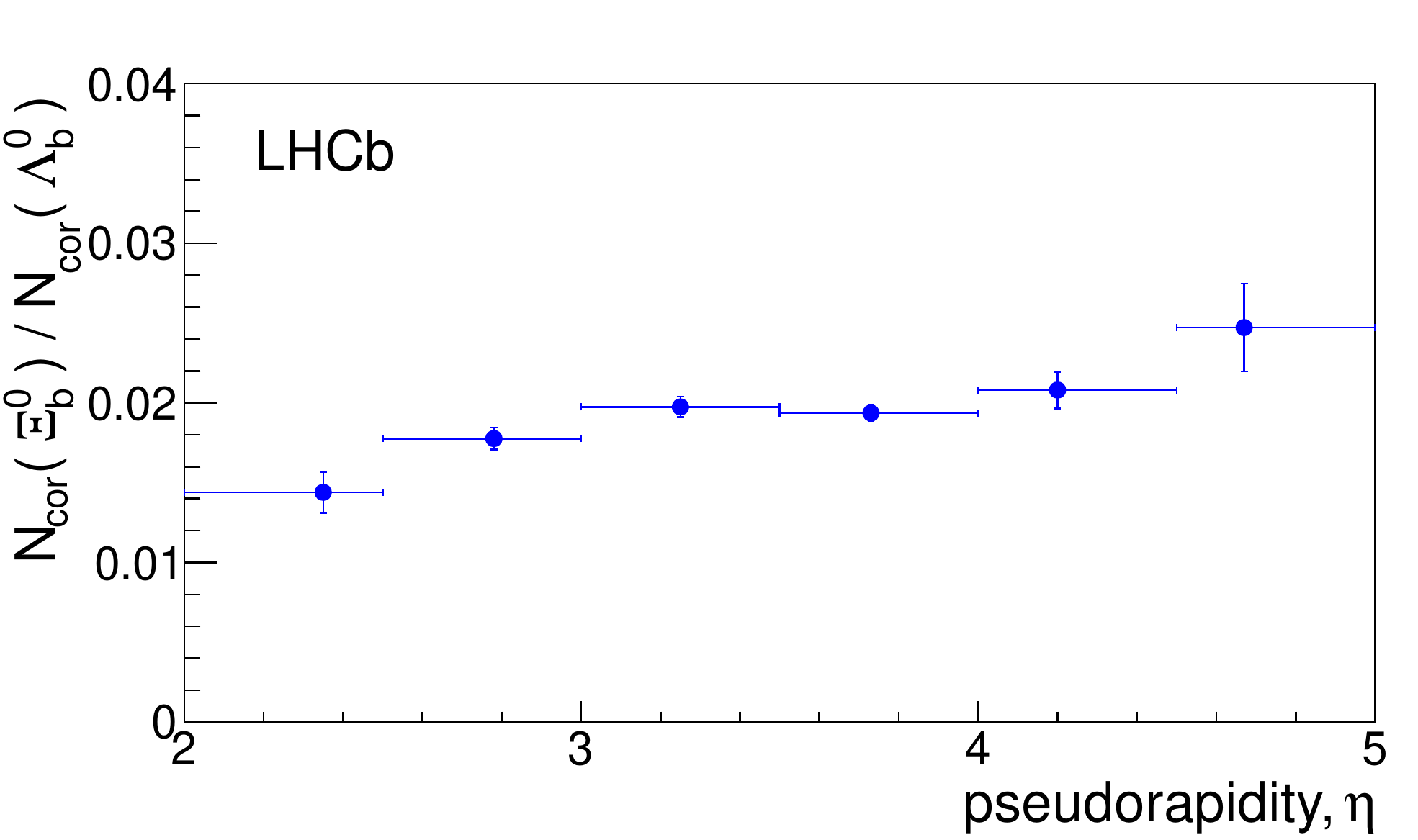}
\caption{\small{Efficiency-corrected yield ratio of $\Xibz\to\Xic\pim$ relative to $\Lb\to\Lc\pim$ decays as functions of
(left) \pt and (right) pseudorapidity, $\eta$. The points are positioned along the horizontal axis at the weighted average 
value within each bin. The uncertainties are statistical only.}}
\label{fig:Xb_kinRatio}
\end{figure}

We have also investigated the relative production rates of  $\Xibz$ and $\Lb$ baryons as
functions of $\pt$ and $\eta$. The $\pt$ bin boundaries are 0, 4, 6, 8, 10, 12, 16, 20, up to a maximum of 
30\gevc, and the $\eta$ bins are each 0.5 units wide ranging from 2 to 5. 
The efficiency-corrected yield ratios are shown in Fig.~\ref{fig:Xb_kinRatio}.
A smooth change in the relative production rates, at about the 10-20\% level, is observed.
Since the \pt dependence of $\Xibz$ and $\Lb$ production
are similar, this implies that the steep \pt dependence of $\Lb$ baryon to $\Bz$ meson production 
measured in Ref.~\cite{LHCb-PAPER-2014-004} also occurs for $\Xibz$ baryons.

\begin{figure}[tb]
\centering
\includegraphics[width=0.48\textwidth]{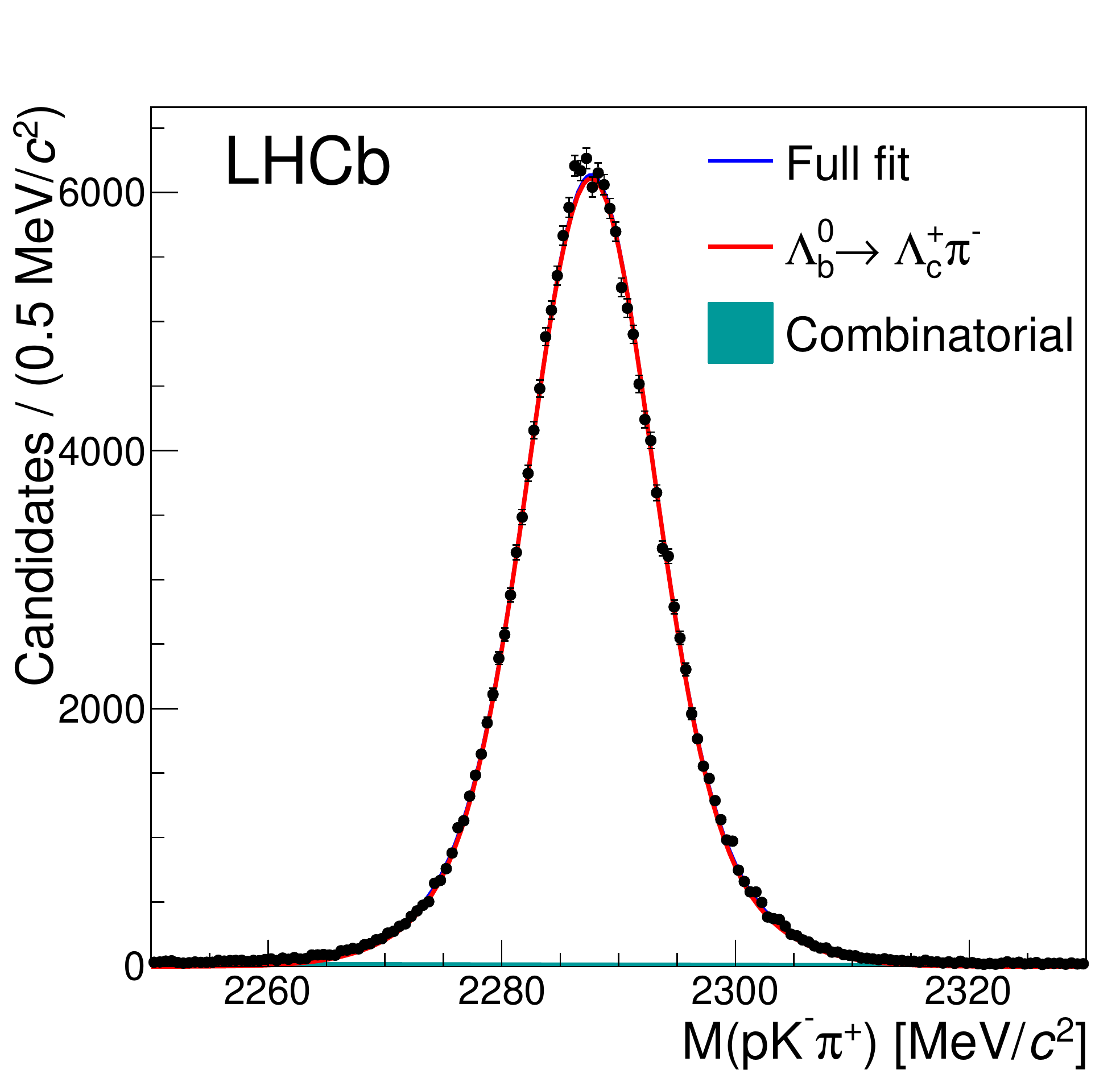}
\includegraphics[width=0.48\textwidth]{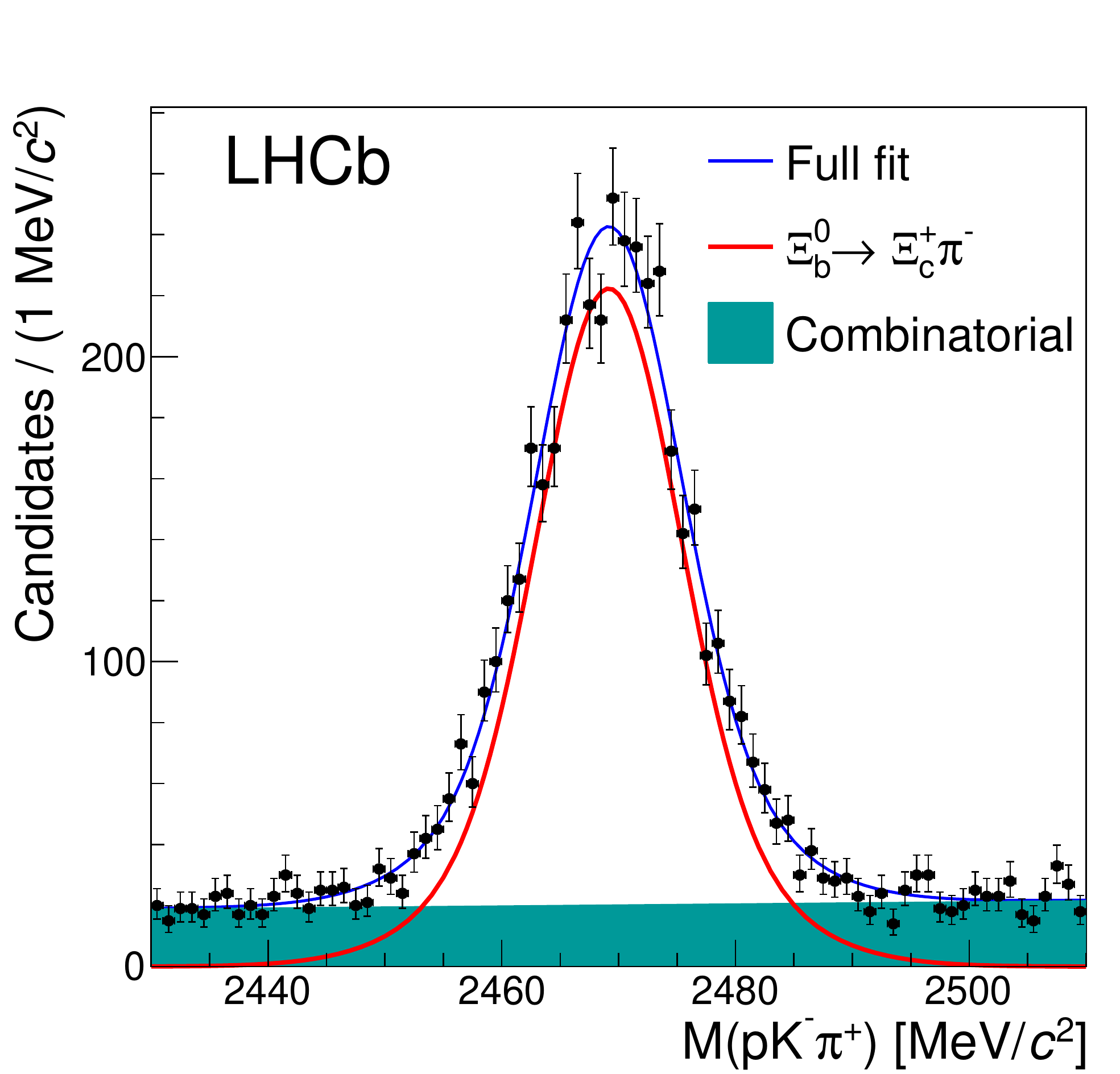}
\caption{\small{Distributions of the $p\Km\pip$ invariant mass for (left) $\Lc$ and (right) $\Xic$ candidates
along with the projections of the fit.}}
\label{fig:xcmass}
\end{figure}

The large sample of $\Xibz\to\Xic\pim$ decays is exploited to measure the $\Xic$ mass. 
Signal $X_b$ candidates within 50\mevcc of their respective peak values are selected, and a simultaneous fit to
the $\Lc$ and $\Xic$ mass spectra is performed. For this measurement, we remove the 20\mevcc restriction on the $X_c$ mass.
The sum of two CB functions is used to describe the signal and an exponential shape describes the background.
The signal shape parameters are common, except for their means and widths. The larger $\Xic$ resolution
is due to the greater energy release in the decay.
The mass distributions and the results of the fit are shown in Fig.~\ref{fig:xcmass}.
\noindent The fitted mass difference is
\begin{align*}
\Delta M_{X_c}\equiv M(\Xic)-M(\Lc) = 181.51\pm0.14\,(\rstat)\mevcc. 
\end{align*}

The results presented are all ratio or difference measurements, reducing their sensitivity to
most potential biases. A summary of the systematic uncertainties is given in Table~\ref{tab:syst}.
Unless otherwise noted, systematic uncertainties are assigned by taking the difference between the nominal result and
the result after a particular variation. In all measurements, possible dependencies on the signal and
background models are investigated by exploring alternative shapes and fit ranges (for mass differences). 
Uncertainties are combined by summing all sources of uncertainty in quadrature.

For the mass difference measurements, common and separate variations in the fraction of $X_b\to X_c\Km$ by $\pm$1\% (absolute) are 
used to assign the cross-feed uncertainty. Shifts in the momentum scale of $\pm$0.03\%~\cite{LHCb-PAPER-2013-011} are applied 
coherently to both signal and normalization mode to determine the momentum scale uncertainty.
Validation of the procedure on simulated decays shows no biases on the results. The uncertainty
due to the limited size of those simulated samples are taken as a systematic error.

For the relative lifetime measurement, 
the relative acceptance uncertainty is dominated by a potential bias in the first time bin. The uncertainty is assessed by
dropping this bin from the fit. Potential bias due to the BDT's usage of $\chisqip$ information is examined
by correcting the data using simulated efficiencies with a tighter BDT requirement. The smaller lifetime of the $\Lb$ baryon
assumed in the simulation (1.426~ps) has a negligible impact on the measured lifetime ratio.
Lastly, the finite size of the simulated samples is also taken into account.
 
For the relative production rate, the signal and background shape uncertainties, and the $X_b\to X_c\Km$ cross-feed 
uncertainties are treated in the same way as above. For the relative acceptance we include contributions from
(i) the geometric acceptance by comparing \pythia 6 and \pythia 8; (ii) the $X_c$ Dalitz structure, by
reweighting the efficiencies according to the distributions seen in data, and (iii) the lower efficiency in the
$0-0.5$~ps bin by requiring $\tau(X_b)>0.5$~\ps. The uncertainty in the relative trigger efficiency is estimated by 
taking the difference in the average trigger efficiency, when using the different TOS/TIS fractions in data and simulation.
A correction and an uncertainty due to the 20\mevcc mass range on $X_c$ is obtained using the results of the $X_c$ mass fits.
The results for the 7\tev and 8\tev data differ by about 1\% and are statistically compatible with each other.
\begin{table*}[tb]
\begin{center}
\caption{\small{Summary of systematic uncertainties on the reported measurements. Below, PR represents the
relative uncertainty on the production ratio measurement.}}
\begin{tabular}{lcccc}
\hline\hline
Source &       $\Delta M_{X_b}$ & $\Delta M_{X_c}$ & $\tau(\Xibz)/\tau(\Lb)$ & PR \\
       &        (\mevcc)        &    (\mevcc)      &      (\%)              & (\%) \\
\hline
Signal \& back. model &   0.06      &       0.05    & 0.1   &  0.5  \\
$X_c\Km$ reflection    &   0.02      &       $-$     &  $-$  &  0.3 \\  
Momentum scale    &   0.06      &       0.06    &  $-$  &   - \\  
Sim. sample size  &  0.14       &       0.07    & 0.9 &  0.6 \\
Detection efficiency  &  $-$        &        $-$    & 0.4 &  1.0 \\  
BDT requirement   &  $-$        &        $-$    & 0.2 &  $-$ \\  
Trigger           &  $-$        &        $-$    & $-$   &  1.3 \\
$X_c$ mass range &  $-$        &        $-$    & $-$   &  0.3 \\   
\hline
Total          &   0.17         &        0.10   & 1.0 &  1.9 \\
\hline\hline
\end{tabular}
\label{tab:syst}
\end{center}
\end{table*}
In summary, a 3\invfb $pp$ collision data set is used to make the first measurement of the $\Xibz$ lifetime. 
The relative and absolute lifetimes are
\begin{align*}
\frac{\tau_{\Xibz}}{\tau_{\Lb}} &= 1.006\pm0.018\,(\rstat)\pm0.010\,(\rsyst),  \\
\tau_{\Xibz} &= 1.477\pm0.026\,(\rstat)\pm0.014\,(\rsyst)\pm0.013\,(\Lb)~{\rm ps},
\end{align*}
\noindent where the last uncertainty in $\tau_{\Xibz}$ is due to the precision of
$\tau_{\Lb}$~\cite{LHCb-PAPER-2014-003}.
This establishes that the $\Xibz$ and $\Lb$ lifetimes are equal to within 2\%.
We also make the most precise measurements of the mass difference and $\Xibz$ mass as
\begin{align*}
M(\Xibz)-M(\Lb) &= 172.44\pm0.39\,(\rstat)\pm0.17\,(\rsyst)\mevcc, \\
M(\Xibz) &= 5791.80\pm0.39\,(\rstat)\pm0.17\,(\rsyst)\pm0.26\,(\Lb)\mevcc,
\end{align*}
\noindent where we have used $M(\Lb) = 5619.36\pm0.26$\mevcc~\cite{LHCb-PAPER-2014-002}.
The mass and mass difference are consistent with, and about five times more 
precise than the value recently obtained in Ref.~\cite{LHCb-PAPER-2013-056}.

We also measure the mass difference $M(\Xic)-M(\Lc)$, and the corresponding $\Xic$ mass, yielding
\begin{align*}
M(\Xic)-M(\Lc) &= 181.51\pm0.14\,(\rstat)\pm0.10\,(\rsyst)\mevcc, \\
M(\Xic) &= 2467.97\pm0.14\,(\rstat)\pm0.10\,(\rsyst)\pm0.14\,(\Lc)\mevcc, 
\end{align*}
\noindent where $M(\Lc)=2286.46\pm0.14$\mevcc~\cite{PDG2012} is used. 
These values are consistent with and at least three times more precise than other measurements~\cite{PDG2012,Aaltonen:2014wfa}.

Furthermore, the relative yield of $\Xibz$ and $\Lb$ baryons as functions of $\pt$ and $\eta$ are
measured, and found to smoothly vary by about 20\%. 
The relative production rate inside the LHCb acceptance is measured to be
\begin{align*}
\frac{f_{\Xibz}}{f_{\Lb}}\cdot\frac{\br(\Xibz\to\Xic\pim)}{\br(\Lb\to\Lc\pim)}\cdot\frac{\br(\Xic\to p\Km\pip)}{\br(\Lc\to p\Km\pip)} = (1.88\pm0.04\pm0.03)\times10^{-2}. 
\end{align*}
\noindent The first fraction is the ratio of fragmentation fractions, $b\to\Xibz$ relative to $b\to\Lb$, and
the remainder are branching fractions. Assuming naive Cabibbo factors~\cite{Cabibbo:1963yz}, namely $\br(\Xibz\to\Xic\pim)/\br(\Lb\to\Lc\pim)\approx1$
and $\br(\Xic\to p\Km\pip)/\br(\Lc\to p\Km\pip)\approx0.1$, one obtains $\frac{f_{\Xibz}}{f_{\Lb}}\approx0.2$.
The results presented in this paper provide stringent tests of models that predict the properties of beauty hadrons.

\section*{Acknowledgements}
\noindent We express our gratitude to our colleagues in the CERN
accelerator departments for the excellent performance of the LHC. We
thank the technical and administrative staff at the LHCb
institutes. We acknowledge support from CERN and from the national
agencies: CAPES, CNPq, FAPERJ and FINEP (Brazil); NSFC (China);
CNRS/IN2P3 and Region Auvergne (France); BMBF, DFG, HGF and MPG
(Germany); SFI (Ireland); INFN (Italy); FOM and NWO (The Netherlands);
SCSR (Poland); MEN/IFA (Romania); MinES, Rosatom, RFBR and NRC
``Kurchatov Institute'' (Russia); MinECo, XuntaGal and GENCAT (Spain);
SNSF and SER (Switzerland); NASU (Ukraine); STFC and the Royal Society (United
Kingdom); NSF (USA). We also acknowledge the support received from EPLANET, 
Marie Curie Actions and the ERC under FP7. 
The Tier1 computing centres are supported by IN2P3 (France), KIT and BMBF (Germany),
INFN (Italy), NWO and SURF (The Netherlands), PIC (Spain), GridPP (United Kingdom).
We are indebted to the communities behind the multiple open source software packages on which we depend.
We are also thankful for the computing resources and the access to software R\&D tools provided by Yandex LLC (Russia).



\addcontentsline{toc}{section}{References}
\setboolean{inbibliography}{true}

\ifx\mcitethebibliography\mciteundefinedmacro
\PackageError{LHCb.bst}{mciteplus.sty has not been loaded}
{This bibstyle requires the use of the mciteplus package.}\fi
\providecommand{\href}[2]{#2}

\end{document}